%
%
%
\def\gap {\vskip 6pt}
\def\indenton {\parindent=20pt}
\def\indentoff {\parindent=0pt}
\def\ket#1{\vert #1\rangle }
\def\bra#1{\langle #1\vert }
\def\braket#1#2{\langle #1 | #2 \rangle }
\def\ref#1#2#3#4#5#6{[#1] #2, {\sl #3}, {\bf #4}, #5, (#6).\vskip 6pt 
\par}
\magnification=\magstep1
\vsize=25 true cm
\goodbreak
\indenton
\vskip 1cm
\centerline {\bf WANNIER FUNCTIONS FOR LATTICES IN A MAGNETIC FIELD II:}
\centerline {\bf EXTENSION TO IRRATIONAL FIELDS}
\vskip 1cm
\centerline {Michael Wilkinson\footnote *{Address after 1st August 2000:
Faculty of Mathematics and Computing, The Open University,
Milton Keynes, MK7 6AA, Bucks, England.}}
\vskip 1cm
\centerline {Department of Physics and Applied Physics,}
\centerline {John Anderson Building,}
\centerline {University of Strathclyde,}
\centerline {Glasgow, G4 0NG,}
\centerline {Scotland, U.K.}
\vskip 1cm
\centerline{\bf Abstract}
\gap
This paper extends earlier work on the definition of 
Wannier functions for Bloch electrons in a magnetic field. 
Extensions to irrational as well as rational 
magnetic fields are defined, and their properties investigated. 
The results are used to give a generalisation of the Peierls 
effective Hamiltonian which is valid when the magnetic flux per 
unit cell is close to any rational number.
\vfill
\eject
%
%
%
%
\noindent{\bf 1. Introduction}
\gap
Wannier functions are localised basis states which span
a band of Bloch eigenfunctions [1]. The use of localised 
basis functions can be convenient both technically and 
conceptually, particularly when considering perturbations
which are themselves spatially localised. There are 
difficulties in defining satisfactory Wannier states
when a magnetic field is applied to the lattice. 
Firstly, the eigenfunctions are typically not Bloch states:
in two dimensions the eigenfunctions are only Bloch states
if the ratio $\beta$ of the flux quantum to the magnetic flux 
per unit cell is a rational number [2,3] (in these cases I 
will write $\beta=p/q$, where $p$ and $q$ are integers with no 
common divisor). Secondly, even when the magnetic field is rational 
in this sense, conventional Wannier states only have satisfactory
localisation properties if a topological invariant 
(the Chern index) chararterising a Bloch band is equal to zero [4,5]. 
In a previous paper (reference [6]) it was shown how this 
latter difficulty could be overcome, for two dimensional 
lattices, in the case where the magentic flux per unit cell 
is rational. In [6] I showed how to obtain a complete set of
states which span a Bloch band, and which retain all of the
useful properties of conventional Wannier functions. 
Two different definitions were examined, termed type I
and type II Wannier functions.
The definition of these states contains the Chern index
$M$ of the band, and in both cases they reduce to the 
conventional Wannier function when $M=0$.

The purpose of the present paper is twofold. The first objective
is to show how the definition of Wannier functions can 
be usefully extended to irrational fields, despite the fact that 
Bloch bands do not exist in this case. Some of the results for
type II Wannier functions are anticipated in earlier papers
by the same author [7], [8] (the latter in collaboration
with R. J. Kay). These earlier papers discussed \lq irrational' 
generalised Wannier functions for the special case of the 
\lq phase space lattice Hamiltonian', a one dimensional
model which reprenents many of the features of Bloch 
electrons in a magnetic field. The form of the Wannier
functions of the full Hamiltonian for irrational
magnetic fields is related to the results for the phase 
space lattice Hamiltonian, but the generalisations are 
not obvious. The derivation given here is also more satisfying in
that it uses only minimal algebraic properties, and that
results are obtained for both types of Wannier functions introduced 
in [6].

The second objective is to use the generalised Wannier 
functions to obtain a very general form of the Peierls
effective Hamiltonian [9,10], in a form suitable
for systematic analysis. For simplicity we consider only
a two-dimensional case where the electron is confined to
a plane, and perturbed by a magnetic field in the perpendicular 
direction (with cartesian coordinate $z$). A comprehensive 
treatment of the three dimensional case introduces the
complication of an additional commensurability parameter,
but is straightforward when the field is aligned
with one of the crystal axes. The Peierls Hamiltonian
is a one-dimensional effective Hamiltonian which 
describes the effect of a uniform magnetic field
perturbing a band of Bloch states. If the dispersion
relation is ${\cal E}(k_x,k_y)$, the Peierls effective
Hamiltonian takes the form 
$$\hat H\sim {\cal E}(\hat K_x,\hat K_y)
\eqno(1.1)$$
where $\hat K_x$ and $\hat K_y$ are generators
of the magnetic translation operators, $\hat T({\bf R})$ (These
are defined in section 3; they were introduced in [11,12], and 
are discussed concisely in [6]). These satisfy
$$[\hat K_x,\hat K_y]={\rm i}
{2\pi \beta^{-1}\over{\vert {\bf A}_1 \wedge {\bf A}_2\vert}}
\eqno(1.2)$$
where ${\bf A}_i$ are the basis vectors for the lattice.
Many derivations of this relationship exist where the
dispersion relation is that of the $B=0$ problem.
This paper considers the case where the dispersion relation
is that of the system with any rational magnetic field $p/q$,
showing that the Peierls effective Hamiltonian is
applicable in this case. It is shown that the commutator 
(1.2) is replaced by one which depends upon the Chern index:
the general form of (1.2) is
$$[\hat K_x,\hat K_y]={\rm i}
{2\pi \gamma\over {\vert{\bf A}_1\wedge {\bf A}_2}\vert}
\eqno(1.3)$$
where $\gamma $ is another dimensionsionless parameter
characterising the magnetic field. The value of $\gamma $ depends 
upon the value of the Chern integer $M$, and
upon another integer $N$ which satisfies
$$qM+pN=1\ .\eqno(1.4)$$
The dimensionless effective magnetic field $\gamma $ is
$$\gamma ={q\beta -p \over{M+N\beta}}\ .\eqno(1.5)$$
The expression (1.5) can be surmised from results obtained previously
for the phase space lattice Hamiltonian [7,8]. The derivation
presented here indicates how the effective Hamiltonian
can be obtained for the full Hamiltonian, rather than
a one-dimensional model. This issue has also been considered
by Chang and Niu [13], who also discussed an heuristic approach
to determining the first order correction to the effective 
Hamiltonian. The method described here allows a systematic
development of the effective Hamiltonian, using similar 
techniques to those applied to the phase space lattice Hamiltonian
in reference [14]. It also has the advantage
that some of the complicated intermediate steps in the algebra
of references [7] and [14] are given a more transparent interpretation. 

Sections 2 and 3 respectively summarise the essential 
definitions and principal results from [6], and a 
representation of the Hamiltonian as a sum of magnetic 
translation operators. The latter will be essential to the 
derivation of the general effective Hamiltonian. 

Section 4 describes the extension of the Wannier 
functions obtained in [6] to irrational magnetic fields, and
a corresponding extension of the definition of Bloch states.
The next four sections consider various properties of the
generalised Wannier functions and Bloch states. Section
5 discusses the effect upon the Wannier functions of a 
transformation of the Bloch states. A 
\lq gauge transformation' of the form
$$\ket {B({\bf k})}\to \ket {B'({\bf k})}
=\exp[{\rm i}\theta ({\bf k})]\ket {B({\bf k})}
\eqno(1.6)$$
is applied to the Bloch states, with $\theta ({\bf k})$ a periodic
function. The Wannier states derived 
from the gauge transformed Bloch states can be obtained from the 
original Bloch states by the action of an operator, which is 
obtained in section 5. 
Similarly, section 6 determines 
an operator acting on the Wannier states which is the 
image of a translation operator acting on the Bloch states.

Section 7 computes the Dirac bracket of two generalised Bloch
states, $\braket{B'({\bf k}')}{B({\bf k})}$, which will
be required for determining matrix elements of the Hamiltonian. 
Section 8 introduces some notational devices which simplify
and illuminate the rather complex expressions obtained 
earlier, showing how they can written in terms of translation
operators with algebra analogous to that of the magnetic 
translation group. Finally, in section 9 these results are used 
to obtain the general form for the Peierls efffective Hamiltonian.
\gap\gap\gap
\vfill
\eject
%
%
%
\noindent{\bf 2. Summary of earlier results}
\gap
The purpose of this section is to present, for the convenience 
of the reader, a summary of some of the principal definitions
and equations from the earlier paper, reference [6]. The lattice vectors
are written ${\bf R}=n_1{\bf A}_1+n_2{\bf A}_2$, and
the reprocal lattice vectors are ${\bf K}=n_1{\bf a}_1+n_2{\bf a}_2$,
with ${\bf a}_i.{\bf A}_j=2\pi \delta_{ij}$.

The magnetic translation operators $\hat T({\bf R})$ 
introduced by Zak [11] and Brown [12] are 
of fundamental importance. They are a representation of
the symmetry of the system: if ${\bf R}$ is a lattice
vector, the $\hat T({\bf R})$ commutes with the Hamiltonian.
The magnetic translation operators do not commute among
themselves, and their composition rule can be written 
in the form
$$\hat T({\bf R}_1)\hat T({\bf R}_2)=
\exp\biggl[{\pi {\rm i}\over \beta}
{({\bf R}_1\wedge {\bf R}_2)\over{({\bf A}_1\wedge {\bf A}_2)}}\biggr]
\hat T({\bf R}_1+{\bf R}_2)
\eqno(2.1)$$
where $\beta $ is the flux quantum divided by the magnetic flux 
per unit cell. The magnetic translation operators are discussed
concisely in [6].

When conventional Wannier functions
are defined, it is assumed that the Bloch states are periodic
functions of the Bloch wavevector ${\bf k}$, as well as being
eigenfunctions of the lattice translation operators 
$\hat T({\bf A}_i)$, with eigenvalues 
$\exp[{\rm i}{\bf k}.{\bf A}_i]$. 
In the case where a rational magnetic field (with $q/p$ flux
quanta per unit cell) is applied, in general both of these conditions
need to be modified. The Bloch states are $p$ fold degenerate,
and their phase increases by $2\pi M$ on traversing the boundary
of the unit cell. Throughout this paper, the following choice
for the eigenvalue and periodicity conditions is preferred:
$$\hat T({\bf A}_1)\ket{B({\bf k})}=
\exp[{\rm i}{\bf k}.{\bf A}_1]\ket{B({\bf k}-q{\bf a}_2/p)}\eqno(2.2a)$$
$$\hat T({\bf A}_2)\ket{B({\bf k})}=
\exp[{\rm i}{\bf k}.{\bf A}_2]\ket{B({\bf k})}\eqno(2.2b)$$
$$\ket{B({\bf k}+{\bf a}_1/p)}=
\exp[{\rm i}M{\bf k}.{\bf A}_2]\ket{B({\bf k})}\eqno(2.2c)$$
$$\ket{B({\bf k}+{\bf a}_2)}=\ket{B({\bf k})}\ .\eqno(2.2d)$$
Bloch states with their phases chosen to satisfy (2.2a)
and (2.2b), and with degenerate states resolved so that
(2.2c) is satisfied will be termed {\sl canonical}
Bloch states.
Except when $p=1$ and $M=0$, these conditions depend
upon the choice of lattice basis vectors ${\bf A}_i$.

The method for constructing the Wannier functions is based upon 
the following observation: if the Bloch states are canonical, the state
$\ket {C({\bf k})}=\hat T(-pM{\bf k}.{\bf A}_2/2\pi)\ket {B({\bf k})}$
is periodic on the Brillouin zone of the superlattice spanned by
$p{\bf A}_1$, ${\bf A}_2$,
and Wannier functions $\ket {\chi ({\bf R})}$ are obtained
by integrating the state $\ket {C({\bf k})}$ with weight 
$\exp[{\rm i}{\bf k}.{\bf R}]$. In the case
of standard Wannier functions, all of the Wannier states
are obtained be applying translation operators to a single
fundamental Wannier state. In the magnetic case, the full
set of Wannier states is obtained by applying lattice translations
to $\vert N \vert $ fundamental type I Wannier states, 
$\ket {\chi_\mu}=\ket {\chi (\mu {\bf A}_1)}, \ \mu=0,..,\vert N\vert -1$,
where $N$ satisfies (1.4). The relation between the
Bloch and type I Wannier states is 
$$\ket{B({\bf k})}=\sum_{{\bf R}=n_1{\bf A}_1+n_2{\bf A}_2}
\exp[-{\rm i}{\bf k}.{\bf R}]\,\hat T(n_2{\bf A}_2)\,\hat T(n_1{\bf A}_1)
\,\hat T \Bigl({pM\over{2\pi}}({\bf k}.{\bf A}_1){\bf A}_2 \Bigr)$$
$$\times \sum_{\mu=0}^{\vert N\vert -1} 
\exp[-{\rm i}p\mu({\bf k}.{\bf A}_1)]\ket{\chi_\mu}
\ .\eqno(2.3)$$
A somewhat more natural representation of the Bloch states
uses an alternative set of fundamental Wannier states: the type
II Wannier states are defined by
$$\ket{\phi_\mu}={1\over N}
\sum_{\mu'=0}^{\vert N\vert -1}\exp[-2\pi{\rm i}\mu \mu'/N]
\hat T(-\mu'{\bf A}_1/N)\ket{\chi_{\mu'}}
\ .\eqno(2.4)$$
One advantage of using the type II Wannier states is that
upon expanding the Bloch states in terms of the $\ket {\phi_\mu}$
states, the summation over $\mu $ no longer depends upon ${\bf k}$: 
the Bloch states are given in terms of the type II states by the 
relation
$$\ket{B({\bf k})}=
\!\!\!\!\!\sum_{{\bf R}=n_1{\bf A}_1/N+n_2{\bf A}_2}\!\!\!
\exp[-{\rm i}{\bf k}.{\bf R}]
\sum_{\mu =0}^{\vert N\vert -1}
\exp[2\pi {\rm i}n_1\mu/N]$$
$$\times \hat T(n_2{\bf A}_2)\hat T(n_1{\bf A}_1/N) 
\hat T \Bigl({pM\over{2\pi}}({\bf k}.{\bf A}_1){\bf A}_2\Bigr)
\ket{\phi_\mu}
\ .\eqno(2.5)$$
The other advantage of the type II Wannier states is
that their transformations under a change of lattice
basis vectors are simpler [6].
\gap\gap\gap
%
%
%
%
\noindent{\bf 3. The Hamiltonian in terms of translation operators}
\gap
Here the objective is to represent the Hamiltonian as 
a sum of magnetic translation operators: this will facilitate 
the construction of the effective Hamiltonian. The Hamiltonian is
$$\hat H={1\over{2m}}\bigl(\hat {\bf p}-e{\bf A}(\hat {\bf r})\bigr)^2
+V(\hat {\bf r})$$
$$V({\bf r})=V({\bf r}+{\bf R}),\ \ \ {\bf R}=n_1{\bf A}_1+n_2{\bf A}_2
\eqno(3.1)$$
with the magnetic field generated by a linear vector potential,
constructed using a matrix $\tilde {\cal B}$ with elements
${\cal B}_{ij}$:
$${\bf A}({\bf r})=\tilde {\cal B}{\bf r},\ \ \ 
\nabla \wedge {\bf A}=B{\bf e}_3,\ \ \ 
{\cal B}_{21}-{\cal B}_{12}=B
\ .\eqno(3.2)$$
The magnetic translations $\hat T({\bf R})$ have a generator
$\hat {\bf P}=\hat P_1{\bf e}_1+\hat P_2{\bf e}_2$:
$$\hat T({\bf R})=\exp[-{\rm i}\hat {\bf P}.{\bf R}/\hbar]\eqno(3.3)$$
$$\hat {\bf P}=\hat {\bf p}-e\tilde {\cal B}^T\hat {\bf r}
\ .\eqno(3.4)$$
It will also be useful to define a set of conjugate generators
$\hat P^\ast_i$:
$$\hat {\bf P}^\ast=\hat {\bf p}-e\tilde {\cal B}\hat {\bf r}
\ .\eqno(3.5)$$
The generators $\hat P_i$, $\hat P^\ast_i$ satisfy the commutation 
relations, where $\varepsilon_{ij}$ is the antisymmetric symbol,
with elements $\varepsilon_{11}=\varepsilon_{22}=0$, 
$\varepsilon_{12}=-\varepsilon_{21}=1$:
$$[\hat P_i,\hat P_j]=-{\rm i}e\hbar B\varepsilon_{ij}\eqno(3.6)$$
$$[\hat P_i^\ast ,\hat P_j^\ast]={\rm i}e\hbar B\varepsilon_{ij}
\eqno(3.7)$$
$$[\hat P_i,\hat P_j\ast]=0
\ .\eqno(3.8)$$
The coordinate vector $\hat {\bf r}$ can be expressed in terms of the 
generators $\hat {\bf P}$ and $\hat {\bf P}^\ast$: from (3.4) and (3.5)
it follows that $\hat P^\ast_i-\hat P_i=eB\varepsilon_{ij}\hat r_j$ 
(where, from here until the end of section 3, 
repeated indices are summed over). This can be 
inverted to give
$$\hat r_i={1\over {eB}}\varepsilon_{ij}(\hat P_j-\hat P^\ast_j)
\ .\eqno(3.9)$$
The Hamiltonian can now be written
$$\hat H={1\over{2m}}\hat P^\ast_i\hat P^\ast_i+\sum_{\bf k}
V_{\bf k}\exp[{\rm i}{\bf k}.\hat {\bf r}]\eqno(3.10)$$
where the ${\bf k}=n_1{\bf a}_1+n_2{\bf a}_2$ are vectors
in the reciprocal lattice, with basis vectors satisfying
${\bf a}_i.{\bf A}_j=2\pi \delta_{ij}$. Expressing the 
$\hat {\bf r}$ using (3.9), and using the fact that
$\hat P_i$ and $\hat P^\ast_j$ commute, (3.10) can
be written in the form
$$\hat H={1\over {2m}}\hat P^\ast_i\hat P^\ast_i
+\sum_{\bf k}V_{\bf k}\ \exp[{\rm i}k_i\varepsilon_{ij}\hat P_j/eB]
\ \exp[-{\rm i}k_i\varepsilon_{ij}\hat P^\ast_j/eB]$$
$$={1\over{2m}}\hat P^\ast_i\hat P^\ast_i+\sum_{\bf k}V_{\bf k}
\ \hat T^\ast(-\hbar {\bf k}^\ast/eB)\ \hat T(\hbar{\bf k}^\ast/eB)
\eqno(3.11)$$
where ${\bf k}^\ast=k^\ast_i{\bf e}_i$ and $\hat T^\ast({\bf R})$
are defined by
$$k_i^\ast=\varepsilon_{ij}k_j,\ \ \ \
\hat T^\ast({\bf R})=\exp[-{\rm i}\hat {\bf P}^\ast.{\bf R}/\hbar]
\ .\eqno(3.12)$$
The Hamiltonian is therefore expressed in terms of a sum
of magnetic translation operators, with operator-valued
coefficients $\hat V_{\bf k}$:
$$\hat H=\sum_{{\bf k}=n_1{\bf a}_1+n_2{\bf a}_2}\hat V_{\bf k}
\ \hat T(\hbar {\bf k}^\ast/eB)\ .\eqno(3.13)$$
The operators $\hat V_{\bf k}$ commute with the magnetic translation
operators, and are given by
$$\hat V_{\bf k}={1\over{2m}}\delta_{{\bf k},{\bf 0}}\,
\hat {\bf P}^\ast.\,\hat {\bf P}^\ast+V_{\bf k}\ 
\hat T^\ast(-\hbar {\bf k}^\ast/eB)
\ .\eqno(3.14)$$
It is desireable to express the vectors ${\bf k}^\ast$ in terms
of the real-space lattice basis vectors ${\bf A}_1$, ${\bf A}_2$.
The vectors corresponding to reciprocal lattice vectors ${\bf a}_i$
are denoted by ${\bf k}_i^\ast$.
Writing ${\bf A}_i=A_{ij}{\bf e}_j$ and 
${\bf a}_i=a_{ij}{\bf e}_j$, the matrices $\tilde A=\{A_{ij}\}$
and $\tilde a=\{a_{ij}\}$ satisfy $\tilde A\tilde a^T=2\pi \tilde I$.
It follows that 
$${\bf k}^\ast_1={2\pi\over {{\rm det}(\tilde A)}}
(-A_{21}{\bf e}_1-A_{22}{\bf e}_2)$$
$${\bf k}^\ast_2={2\pi\over {{\rm det}(\tilde A)}}
(A_{11}{\bf e}_1+A_{12}{\bf e}_2)
\ .\eqno(3.15)$$
Noting that the ${\rm det}(\tilde A)$ is equal to the area
${\cal A}$ of the unit cell, the Hamiltonian (3.13) can then 
written
$$\hat H=\sum_{n_1=-\infty}^\infty \sum_{n_2=-\infty}^\infty
\hat V_{\bf k}\ 
\hat T\biggl({h\over{eB{\cal A}}}(n_2{\bf A}_1-n_1{\bf A}_2)\biggr)
\eqno(3.16)$$
where ${\bf k}=n_i{\bf a}_i$.
The elementary transformations associated with this representation
of the Hamiltonian, $\hat T_1=\hat T(-h{\bf A}_2/eB{\cal A})$ and
$\hat T_2=\hat T(h{\bf A}_1/eB{\cal A})$ therefore span a lattice
which is aligned with the crystal lattice, but scaled by a 
dimensionless factor
$$\beta ={h\over{eB{\cal A}}}
={{\rm flux\ quantum}\over{{\rm flux\ per\ unit\ cell}}}
\ .\eqno(3.17)$$
\gap\gap\gap
%
%
%
%
\noindent{\bf 4. Extension to irrational magnetic fields}
\gap
When the number of flux quanta per unit cell is rational,
the spectrum consists of Bloch bands for which Wannier 
functions have been defined. When the number of flux quanta 
per unit cell is irrational, there are no Bloch bands and
the spectrum is a Cantor set. It is however still possible
to define useful sets of generalised Bloch states and 
corresponding Wannier functions.

The expression giving the Bloch states in terms of the 
type II Wannier states will be generalised, by writing
$$\ket {B({\bf k})}=\sum_{{\bf R}=n_1{\bf A}_1/N+n_2{\bf A}_2}
\exp[-{\rm i}{\bf k}.{\bf R}]\sum_{\mu=0}^{\vert N\vert -1}
\exp[2\pi {\rm i}n_1\mu /N]$$
$$\times\hat T(n_2{\bf A}_2)\,
\hat T(n_1{\bf A}_1/N)\,\hat T(M({\bf k}.{\bf A}_1){\bf A}_2/\kappa)
\,\ket {\phi_\mu}
\eqno(4.1)$$
Straightforward application of the composition law (2.1) for magnetic
translations to the form (4.1) shows that the 
generalised Bloch states satisfy a periodicity condition
$$\ket {B({\bf k}+\kappa{\bf a}_1/2\pi)}=
\exp[{\rm i}M({\bf k}.{\bf A}_2)]\ket {B({\bf k})}\eqno(4.2)$$
provided that $\exp[2\pi {\rm i}Mn/\beta N]\exp[-{\rm i}\kappa n/N]=1$ 
for all integer $n$. The latter condition is used to determine
allowed values for the constant $\kappa $: this quantity must
satisfy $\kappa\beta =2\pi (M+\beta NJ)$ with $J$ 
an integer. Equation (4.2) is a natural generalisation
of the periodicity condition (2.2c). It is desireable to
define the generalised Bloch states so that as $\beta \to p/q$
they converge to the Bloch eigenstates of the rational case
with $\beta =p/q$. Setting $J=1$ (and using (1.4)), $\kappa$
aproaches $2\pi /p$ as $\beta \to p/q$, which is consistent 
with (2.2c). The appropriate
choice of the constant $\kappa $ defining the dimension of the
Brillouin zone is therefore 
$$\kappa \beta =2\pi (M+\beta N)
\ .\eqno(4.3)$$
Systematic application of (2.1) shows that the states (4.1)
also satisfy other conditions analogous to the standard
Bloch states: collecting together the periodicity properties
and the equations defining the effect of lattice vector translations,
the generalised Bloch states satisfy the relations
$$\ket{B({\bf k}+\kappa {\bf a}_1/2\pi)}
=\exp[{\rm i}M({\bf k}.{\bf A}_2)]\ket{B({\bf k})}\eqno(4.4a)$$
$$\ket{B({\bf k}+{\bf a}_2)}=\ket{B({\bf k})}\eqno(4.4b)$$
$$\hat T({\bf A}_1)\ket{B({\bf k})}=\exp[{\rm i}({\bf k}.{\bf A}_1)]
\ket{B({\bf k}-{\bf a}_2/\beta)}\eqno(4.4c)$$
$$\hat T({\bf A}_2)\ket{B({\bf k})}
=\exp[{\rm i}({\bf k}.{\bf A}_2)]\ket{B({\bf k})}
\ .\eqno(4.4d)$$
\par
Equation (4.1) defined the generalised Bloch states in terms
of type II Wannier states. Using the relation between the type
I and type II Wannier functions given by (2.4), the corresponding
relation giving the generalised Bloch states in terms of
type I Wannier functions is
$$\ket {B({\bf k})}=\sum_{{\bf R}=n_1{\bf A}_1+n_2{\bf A}_2}
\exp[-{\rm i}{\bf k}.{\bf R}]
\,\hat T(n_2{\bf A}_2)\,\hat T(n_1{\bf A}_1)$$
$$\times \hat T(M({\bf k}.{\bf A}_1){\bf A}_2/\kappa )
\sum_{\mu=0}^{\vert N\vert -1}
\exp[-2\pi{\rm i}\mu ({\bf k}.{\bf A}_1)/\kappa]\ket {\chi_\mu}
\eqno(4.5)$$
On systematic application of (2.1) and (4.3), it is found
that the states (4.5) satisfy the canonical Bloch state 
relations in the form (4.4a-d), for any states $\ket {\chi_\mu}$.
The relation between the type I and type II Wannier
functions therefore remains valid in the irrational case.

The generalised Bloch states lie in a Brillouin zone spanned
by the reciprocal lattice vectors $\kappa {\bf a}_1/2\pi$
and ${\bf a}_2$, with area 
${\cal A}_{\bf k}=\kappa \vert {\bf a}_1\wedge {\bf a}_2 \vert/2\pi$.
Applying Born-von Karman boundary conditions, the density of states 
per unit area associated with the set of generalised Bloch states is 
${\cal A}_{\bf k}/4\pi^2$. The area of the real space unit cell,
${\cal A}=\vert {\bf A}_1\wedge {\bf A}_2\vert$ is equal to 
$4\pi^2/\vert {\bf a}_1 \wedge {\bf a}_2\vert$. 
The density of generalised Bloch states states per unit area 
is therefore
$${\cal N}={\kappa \over{2\pi {\cal A}}}
\ .\eqno(4.6)$$
It will now be shown that this density of states  
is precisely what is required for them to form a 
complete set of states for a region of the spectrum bounded
by two gaps.
St\v reda [15] showed that the density of bulk states per unit area for
region of the spectrum bounded by two gaps satisfies
is related to the Hall coefficient $\sigma_{xy}$:
$$\sigma_{xy}=e{\partial {\cal N}\over {\partial B}}
\ .\eqno(4.7)$$
The Hall coefficient is quantised in units of $e^2/h$, and
the Chern number $M$ is the quantum number [5]:
$$\sigma_{xy}=M{e^2\over h}\ .\eqno(4.8)$$
The density of states is clearly correct in the rational case.
Using (4.3) with the relation $\beta =h/eB{\cal A}$ to differentiate (4.6) 
with respect to $B$, equation (4.7) reproduces (4.8). 
This shows that the variation
of the density of generalised Bloch states with respect to magnetic 
field is precisely the same as that of the eigenstates. The
generalised Bloch states are therefore a complete set provided
that they are not linearly related.
\gap\gap\gap
%
%
%
%
\noindent{\bf 5. Images of gauge transformations}
\gap
The gauge transformations considered are of the form (1.6), in which 
the Bloch states
are multiplied by a factor $\exp[{\rm i}\theta ({\bf k})]$.
The cases of rational and irrational fields will be considered
separately. 
\gap
\noindent{\sl 5.1 Rational case}
\gap
In the rational case $\theta $ satisfies
$$\theta ({\bf k}+{\bf a}_1/p)-2\pi L_1=\theta ({\bf k})
=\theta ({\bf k}+{\bf a}_2/p)-2\pi L_2\eqno(5.1)$$
with $L_1$ and $L_2$ integers, so that the gauge transformation
leaves the Bloch states in canonical form. Wannier
functions may be defined for the gauge transformed states.
These Wannier functions will be different from the original
ones, and it is interesting to determine how the transformed
Wannier functions may be obtained from the original ones directly.

The calculation will be presented for the special case where
$$\exp[{\rm i}\theta ({\bf k})]=\exp[{\rm i}{\bf k}.{\bf R}^\ast]
\eqno(5.2)$$
where 
$${\bf R}^\ast=p(L_1{\bf A}_1+L_2{\bf A}_2)\eqno(5.3)$$
is a superlattice vector. More general transformations of the form 
$\theta({\bf k})={\bf k}.{\bf R}^\ast+\epsilon \tilde \theta ({\bf k})$,
with $\tilde \theta({\bf k})$ periodic in $k_1$ and $k_2$, with
period $2\pi/p$, can also be treated for $\epsilon \ll 1$ by
Fourier expanding $\tilde \theta ({\bf k})$. 
The type I Wannier functions of the gauge transformed Bloch states are
$$\ket {\chi'({\bf R})}={p\over {4\pi^2}}\int_{\rm BZ} d{\bf k}\ 
\exp[{\rm i}{\bf k}.({\bf R}+{\bf R}^\ast)]
\hat T\bigl({-pM\over{2\pi}}({\bf k}.{\bf A}_1){\bf A}_2\bigr)
\ket {B({\bf k})}$$
$$=\ket {\chi ({\bf R}+{\bf R}^\ast)}
\ .\eqno(5.4)$$
The fundamental type I Wannier functions, $\ket {\chi_\mu}$, are
a subset of the full set of Wannier states $\ket {\chi ({\bf R})}$,
defined by $\ket {\chi_\mu}=\ket {\chi (p\mu {\bf A}_1)}$: previously
the index $\mu $ was restricted to the range 
$\mu \in \{0,..,\vert N\vert -1\}$ but it is convenient to extend the 
definition by allowing $\mu $ to take any integer value.
The states $\ket {\chi ({\bf R})}$ 
are obtained from the fundamental Wannier functions by the
relation [6] 
$$\ket {\chi(p(Nn_1+\mu){\bf A}_1+n_2{\bf A}_2)}
=\hat T(n_2{\bf A}_2)\,\hat T(n_1{\bf A}_1)\,
\ket {\chi_\mu}
\ .\eqno(5.5)$$
It follows that the extended set of fundamental Wannier 
states satisfies
$$\ket {\chi_{\mu +N}}=\hat T({\bf A}_1)\,\ket {\chi_\mu}
\ .\eqno(5.6)$$
The transformation of the fundamental type I Wannier functions
is therefore
$$\ket {\chi'_\mu}=\hat T(pL_2{\bf A}_2)\,
\ket {\chi_{\mu+L_1}}
\ .\eqno(5.7)$$
\par
The corresponding transformation of the type II Wannier states
is obtained using (2.4) and its inverse relation as follows
$$\ket {\phi'_\mu}={1\over N}\sum_{\mu'=0}^{\vert N\vert -1}
\exp[-2\pi {\rm i}\mu \mu'/N]\hat T(-\mu' {\bf A}_1/N)
\ket {\chi'_{\mu'}}$$
$$={1\over N}\sum_{\mu'=0}^{\vert N\vert -1}
\sum_{\lambda =0}^{\vert N\vert -1}
\exp[2\pi {\rm i}(\lambda-\mu)\mu'/N]\exp[2\pi {\rm i}\lambda L_1/N]$$
$$\times\hat T(-\mu'{\bf A}_1/N)\hat T(pL_2{\bf A}_2)
\hat T((\mu'+L_1){\bf A}_1/N)\ket {\phi_\lambda}
\ .\eqno(5.8)$$
After combining the translation operators, the summations can
be performed: only the term $\lambda =\mu +qL_2 $ contributes,
giving the result
$$\ket {\phi'_\mu}=\exp\biggl[{2\pi {\rm i}(\mu +L_2 q)L_1\over N}\biggr]
\hat T(pL_2{\bf A}_2)\hat T(L_1{\bf A}_1/N)
\ket {\phi_{\mu+qL_2}}$$
$$=\exp\biggl[{2\pi {\rm i}(\mu+{\textstyle{1\over 2}}qL_2)L_1\over N}\biggr]
\hat T\bigl(L_1{\bf A}_1/N+pL_2{\bf A}_2\bigr)\ket {\phi_{\mu+qL_2}}
\ .\eqno(5.9)$$
This expression will be re-cast into a more transparent form
in section 8.
\gap
\noindent{\sl 5.2 Irrational case}
\gap
Now consider the case of gauge transformations of the generalised
Bloch states defined for irrational fields. In order to 
define a transformation of the Wannier
functions, the gauge transformation must leave the Bloch states
in canonical form. If $\beta $ is irrational, equations (4.4b) and
(4.4c) imply that a suitable gauge transformation cannot depend
upon $k_2$. Linear gauge transformations analogous to (5.2) 
are therefore restricted to being of the form
$$\ket{B'({\bf k})}=\exp[2\pi{\rm i}k_1L_1/\kappa]
\ket{B({\bf k})}\ .\eqno(5.10)$$
Using (4.1), a Bloch state may be written in terms
of the Wannier functions $\ket{\phi_\mu '}$ as follows
$$\ket{B'({\bf k})}=
\sum_{n_1=-\infty}^\infty\sum_{n_2=-\infty}^\infty
\sum_{\mu=0}^{\vert N\vert -1}\exp[-{\rm i}k_1(n_1-L_1)/N]
\exp[-{\rm i}k_2n_2]\exp[2\pi {\rm i}\mu (n_1-L_1)/N]$$
$$\times \hat T(n_2{\bf A}_2)\hat T((n_1-L_1){\bf A}_1/N)
\hat T(Mk_1{\bf A}_2/\kappa)\,\ket {\phi'_\mu}
\ .\eqno(5.11)$$
If the Wannier functions generating this state are
$$\ket{\phi'_\mu}=
\exp[2\pi {\rm i}\mu L_1/N]
\hat T(L_1{\bf A}_1/N)\,\ket{\phi_\mu}
\eqno(5.12)$$
then (using (4.3)) it can be seen that $\ket {B'({\bf k})}$
is related to the original Bloch state by (5.10).
This result reduces to a special case of (5.9)
in the case where $\beta =p/q$.
\gap\gap\gap
%
%
%
%
\noindent{\bf 6. Images of translation operators acting upon Wannier states}
\gap
This section discusses the states 
$$\hat T({\bf r})\ket{B({\bf k})}\ ,\ \ \ 
{\bf r}=\beta(\nu_1 {\bf A}_1+\nu_2{\bf A}_2)
\eqno(6.1)$$
with $\nu_1$, $\nu_2$ taking integer values.
It will be demonstrated that they are generalised Bloch
states of the form (4.1), generated by a set of Wannier functions 
$\ket{\phi'_\mu},\ \mu=0,..,\vert N\vert-1$. The transformation
giving these Wannier states in terms the states $\ket{\phi_\mu}$
which generate the original Bloch state $\ket {B({\bf k})}$ will 
be determined. This transformation may be regarded as the image
of the operator $\hat T({\bf r})$ acting on the Wannier functions.

The wavevector ${\bf k}=(k_1,k_2)$ of the state (6.1) is shifted
to $(k_1+\Delta k_1,k_2)$, with $\Delta k_1$ to be determined.
Commuting the operator $\hat T({\bf r})$ to the right using 
(2.1) gives
$$\hat T({\bf r})\ket {B({\bf k})}
=\sum_{n_1=-\infty}^\infty 
\sum_{n_2=-\infty}^\infty \sum_{\mu =0}^{\vert N\vert -1}
\exp[-{\rm i}(k_1+\Delta k_1) n_1/N]\exp[-{\rm i}k_2 n_2]
\exp[2\pi {\rm i}\mu n_1/N]$$
$$\times \hat T(n_2{\bf A}_2)\hat T(n_1{\bf A}_1/N)
\hat T(M(k_1+\Delta k_1){\bf A}_2/\kappa)$$
$$\exp[{\rm i}\Delta k_1n_1/N]\exp[-2\pi{\rm i}\nu_2n_1/N]
\exp[2\pi{\rm i}(k_1+{\textstyle{1\over 2}}\Delta k_1M)\nu_1/\kappa]$$
$$\times 
\hat T(\beta \nu_1{\bf A}_1+(\beta \nu_2-M\Delta k_1/\kappa){\bf A}_2)
\ket {\phi_\mu}
\ .\eqno(6.2)$$
This state is a generalised Bloch state if the product of the
final two phase factors containing $n_1$ is unity. This occurs
if $\Delta k_1=2\pi \nu_2$. In this case the argument of the
last translation operator simplifies, the multiplier of ${\bf A}_2$
becoming $2\pi N\nu_2\beta/\kappa$. The state (6.1) is then
in the form
$$\hat T({\bf r})\ket {B({\bf k})}=\exp[{\rm i}\theta({\bf k}')]
\ket{B'({\bf k}')}
\eqno(6.3)$$
where $\ket {B'({\bf k})}$ is a generalised Bloch state with 
Wannier functions $\ket {\phi'_\mu}$, ${\bf k}'={\bf k}+\nu_2{\bf a}_1$,
and
$$\theta({\bf k})={2\pi k_1M\nu_1\over \kappa}
\ .\eqno(6.4)$$
The Wannier functions generating $\ket {B'({\bf k})}$ are
$$\ket {\phi'_\mu}=\exp[-2\pi^2{\rm i}M\nu_1\nu_2/\kappa]
\hat T(\beta \nu_1{\bf A}_1+2\pi N\beta\nu_2{\bf A}_2/\kappa)
\ket {\phi_\mu}
\ .\eqno(6.5)$$
The phase factor in (6.4) represents a gauge transformation
of the type (5.10). Using (5.12), we may therfore write
$$\hat T({\bf r})\ket {B({\bf k})}=\ket {B''({\bf k}+\nu_2{\bf a}_1)}
\eqno(6.6)$$
where the Wannier states generating $\ket{B({\bf k})}$ are
$$\ket {\phi''_\mu}=\exp[2\pi {\rm i}M\mu \nu_1/N]
\hat T(M\nu_1{\bf A}_1/N)\ket{\phi'_\mu}
\ .\eqno(6.7)$$
The Wannier functions generating the Bloch states 
$\ket {B''({\bf k}+\nu_2{\bf a}_1)}=\hat T({\bf r})\ket {B({\bf k})}$
can now be expressed in terms of the original Wannier states:
$$\ket {\phi''_\mu}=\exp\biggl[{2\pi{\rm i}M\mu \nu_1\over{N}}\biggr]
\hat T\biggl({\kappa\beta\over {2\pi N}}\nu_1{\bf A}_1+
{2\pi N\beta\over \kappa}\nu_2{\bf A}_2\biggr)\,\ket {\phi_\mu}
\ .\eqno(6.8)$$
\gap\gap\gap
%
%
%
%
\noindent{\bf 7. Dirac brackets of generalised Bloch states}
\gap
The objective is to evaluate the matrix element
$$I({\bf k},{\bf k}')
=\braket {B'({\bf k}')}{B({\bf k})}\eqno(7.1)$$
where the $\ket{B({\bf k})}$ and $\ket {B'({\bf k})}$ are 
different generalised Bloch states for irrational magnetic fields.
These Bloch states are 
generated by different type II Wannier states $\ket {\phi_\mu}$ 
and $\ket{\phi'_\mu}$ respctively, using the expansion (4.1).
The resulting expression will later be used to calculate matrix
elements of the form 
$\bra {B({\bf k}')}\hat T({\bf r})\ket {B({\bf k})}$, 
(where ${\bf r}=\nu_1{\bf a}_1+\nu_2{\bf a}_2$),
and
hence matrix elements of the Hamiltonian, by writing 
$\ket {B'({\bf k}+\nu_2{\bf a}_1)}=\hat T({\bf r})\ket {B({\bf k})}$.

Using (4.1) and (2.1), and writing $k_i={\bf k}.{\bf A}_i$, the 
Dirac bracket is
$$I=\sum_{n_1=-\infty}^\infty \sum_{n_1'=-\infty}^\infty
\sum_{n_2=-\infty}^\infty \sum_{n_2'=-\infty}^\infty 
\sum_{\mu=0}^{\vert N\vert -1}\sum_{\mu'=0}^{\vert N\vert -1}
\exp\biggl[{\rm i}(k'_2-k_2)\biggl({n_2+n_2'\over{2}}\biggr)\biggr]$$
$$\times \exp\biggl[{\rm i}
\bigl((k'_1-k_1)+2\pi(\mu-\mu')-{2\pi\over{\beta}}(n_2-n'_2)\bigr)
\biggl({n_1+n_1'\over{2N}}\biggr)\biggr]
\exp\biggl[\biggl({k_1+k_1'\over{2\kappa}}\biggr)(n_1'-n_1)\biggr]$$
$$\times
\exp\biggl[{2\pi{\rm i}\over{N}}\biggl({\mu+\mu'\over{2}}\biggr)
(n_1-n'_1)\biggr]
\exp\biggl[{\rm i}\biggl({k_2'+k_2\over{2}}\biggr)(n_2-n_2')\biggr]$$
$$\times\bra{\phi'_{\mu'}}\hat T\biggl({n_1-n_1'\over{N}}{\bf A}_1
+\bigl(n_2-n_2'+{M\over \kappa}(k_1-k_1')\bigr){\bf A}_2\biggr)
\ket{\phi_\mu}
\ .\eqno(7.2)$$
It is convenient to make changes of variable
$$\eqalign{j&=n_1-n_1'\ \ \ J={n_1+n_1'\over 2}\cr
l&=n_2-n_2'\ \ \ L={n_2+n_2'\over 2}
\ .\cr}
\eqno(7.3)$$
The summations in (7.2) will then run over integer values
of $L$ for even $l$, and over integer plus one half values
of $L$ for odd $l$, similarly for $J$ and $j$. These sums
are most conveneiently evaluated by decomposing them into
four summations:
$$I=\sum_j\sum_l\sum_J\sum_L F(j,l,J,L)$$
$$=\sum_{n=-\infty}^\infty \sum_{n'=-\infty}^\infty
\sum_{m=-\infty}^\infty \sum_{m'=-\infty}^\infty
\biggl[ F(2n,2m,n',m')+F(2n+1,m,n'+{\textstyle{1\over 2}},m')$$
$$+F(2n,2m+1,n',m'+{\textstyle{1\over 2}})
+F(2n+1,2m+1,n'+{\textstyle{1\over 2}},m'+{\textstyle{1\over 2}})
\biggr]
\ .\eqno(7.4)$$
The function $F(j,l,J,L)$ is of the form
$$F(j,l,J,L)=\exp[{\rm i}\alpha_1(l)J]\, \exp[{\rm i}\alpha_2 L]\, 
f(j,l)
\eqno(7.5)$$
where
$$\eqalign{\alpha_1(l)&={1\over N}\bigl[k_1'-k_1+2\pi(\mu-\mu')
-{2\pi\over{\beta}}l\bigr]\cr
\alpha_2&=k_2'-k_2\cr}
\eqno(7.6)$$
and
$$f(j,l)=\sum_{\mu=0}^{\vert N\vert -1}\sum_{\mu'=0}^{\vert N\vert -1}
\exp\biggl[-2\pi {\rm i}\biggl({k_1+k_1'\over{2\kappa}}
-{\mu+\mu'\over{2N}}\biggr)j\biggr]
\exp\biggl[{\rm i}\biggl({k_2+k_2'\over{2}}\biggr)l\biggr]$$
$$\times\bra{\phi'_{\mu'}}\hat T \bigl( {j\over N}{\bf A}_1+
(l+{M\over \kappa}(k_1-k_1')){\bf A}_2\bigr)\ket{\phi_\mu}
\ .\eqno(7.7)$$
Using (7.5), it is seen that the sums over $J$ and $L$ are easily evaluated
using the Poisson summation formula in the form
$$\sum_{n=-\infty}^\infty \exp({\rm i}\alpha n)=2\pi \sum_{m=-\infty}^\infty
\delta (\alpha-2\pi m)
\ .\eqno(7.8)$$
Using this formula,
$$I({\bf k}',{\bf k})={4\pi^2\over N}
\sum_{N_1=-\infty}^\infty \sum_{N_2=-\infty}^\infty
\sum_{n_1=-\infty}^\infty \sum_{n_2=-\infty}^\infty
(-1)^{(n_1N_1+n_2N_2)}\delta(k_2-k_2'-2\pi N_2)$$
$$\times
\sum_{\mu=0}^{\vert N\vert -1}\sum_{\mu'=0}^{\vert N\vert -1}
\delta \bigr(k_1-k_1'-2\pi(\mu-\mu')+{2\pi\over \beta}n_2-2\pi N N_1\bigr)
\exp\biggl[{\rm i}\biggl({k_2+k_2'\over{2}}\biggr)n_2\biggr]$$
$$\times \exp\biggl[-2\pi{\rm i}\biggl({k_1+k_1'\over{2\kappa}}
-{\mu+\mu'\over{2}}\biggr)n_1\biggr]
\bra{\phi'_{\mu'}}\hat T \bigl({n_1\over N}{\bf A}_1+
(n_2+{M\over \kappa}(k_1-k_1')){\bf A}_2\bigr)\ket{\phi_\mu}
\ .\eqno(7.9)$$
Writing
$$\Delta k=2\pi \bigl(q-{p\over \beta}\bigr)\eqno(7.10)$$
and recalling (4.3), the values of $k_1-k_1'$ for which the
matrix element is non-zero may be written in two alternative
forms:
$$k_1-k_1'=l_1\Delta k+l_2\kappa
=2\pi \bigl(L_1+{1\over \beta}L_2\bigr)
\eqno(7.11)$$
where $l_1$, $l_2$ and $L_1$, $L_2$ are all integers. 
The argument of the second delta function in (7.9) can therefore 
be written in terms of $\Delta k$ and $\kappa$. Noting that
$${\partial (L_1,L_2)\over{\partial (l_1,l_2)}}
=\bigg\vert \matrix {q&N\cr -p&M\cr}\bigg\vert=1
\eqno(7.12)$$
it is seen that the sums over $N_1$, $\mu'$ and $n_2$ in (7.9)
may be replaced by a sum over the indices $l_1$, $l_2$
in (7.11). In terms of the new indices $l_1$, $l_2$:
$$\eqalign{
\mu&=\mu'+ql_1-\lambda N,\ \ \ \lambda={\rm int}[(\mu'+ql_1)/N]\cr
N_1&=l_2+\lambda\cr
n_2&=pl_1-Ml_2\ .\cr}
\eqno(7.13)$$
Also, the argument of the translation operator in (7.9) simplifies,
since (using (7.10), (4.3) and (7.11))
$$n_2+{M\over \kappa}(k_1-k_1')
=(pl_1-Ml_2)+{M\over \kappa}(\Delta kl_1+\kappa l_2)$$
$$=\biggl({M\Delta k\over \kappa}+p\biggr)l_1={2\pi \over \kappa}l_1
\ .\eqno(7.14)$$
After renaming some of the dummy indices,
the Dirac bracket may be written in the form
$$I({\bf k}',{\bf k})={4\pi^2\over{N}}
\sum_{N_1=-\infty}^\infty \sum_{N_2=-\infty}^\infty 
\sum_{n_1=-\infty}^\infty \sum_{n_2=-\infty}^\infty
(-1)^{N_2(pn_1-MN_1)+N_1n_2}\, \delta (k_2'-k_2-2\pi N_2)$$
$$\times 
\delta(k_1-k_1'-N_1\kappa -n_1\Delta k)
\exp\biggl[{\rm i}\biggl({k_2+k_2'\over{2}}\biggr)(pn_1-MN_1)\biggr]$$
$$\times \sum_{\mu=0}^{\vert N\vert -1}
\exp\biggl[-2\pi {\rm i}\biggl({k_1+k_1'\over{2\kappa}}-
{\mu+\mu'\over{2}}\biggr)n_2\biggr]
\bra{\phi_{\mu}}\hat T\bigl({n_2\over N}{\bf A}_1
+{2\pi\over \kappa}n_1{\bf A}_2\bigr)\ket{\phi_{\mu+qn_1+\lambda N}}
\ .\eqno(7.15)$$
Using the fact that the type II Wannier functions satisfy 
$\ket {\phi_{\mu +N}}=\ket {\phi_\mu}$, the Dirac bracket (7.1) may 
finally be written in terms of a set of coefficients $I_{n_1n_2}$ in the form
$$I({\bf k}',{\bf k})={4\pi^2\over {N}}
\sum_{N_1=-\infty}^\infty \sum_{N_2=-\infty}^\infty
\sum_{n_1=-\infty}^\infty \sum_{n_2=-\infty}^\infty
(-1)^{N_1n_2+pN_2n_1+MN_1N_2}\,\delta(k_2-k_2'-2\pi N_2)$$
$$\times 
\delta (k_1-k_1'-N_1\kappa-n_1\Delta k)
\exp\biggl[{\rm i}\biggl({k_2+k_2'\over{2}}\biggr)(pn_1-MN_1)\biggr]$$
$$\times 
\exp\biggl[-2\pi {\rm i}\biggl({k_1+k_1'\over{2\kappa}}\biggr)n_2\biggr]
I_{n_1n_2}
\ .\eqno(7.16)$$
The coefficients $I_{n_1n_2}$ are given by
$$I_{n_1n_2}=\sum_{\mu=0}^{\vert N\vert -1}
\exp\biggl[{2\pi {\rm i}\over {N}}
\bigl(\mu+{\textstyle{1\over 2}}qn_1\bigr)n_2\biggr]
\bra{\phi'_\mu}
\hat T\bigl({n_2\over N}{\bf A}_1+{2\pi\over\kappa}n_1{\bf A}_2\bigr)
\ket{\phi_{\mu+qn_1}}
\ .\eqno(7.17)$$
\gap\gap\gap
\noindent {\bf 8. Representations in terms of translation operators}
\gap
The expression (7.17) for the coefficients defining the Dirac bracket,
and the expression (6.8) for the Wannier function image of a 
translation operator acting upon a Bloch state, can both be
expressed more elegantly by defining extensions of the
magnetic translation group.

First I will define a translation operator which acts upon the labels
of the type II Wannier states. For integer values of $\lambda_1$
and $\lambda_2$ they are defined by
$$\hat t(n_1,n_2) \ket{\phi_\mu}
=\exp\biggl[{2\pi {\rm i}M\over{N}}
\bigl(\mu-{\textstyle{1\over 2}}n_1\bigr)n_2\biggr]
\ket{\phi_{\mu-n_1}}
\ .\eqno(8.1)$$
These operators were originally introduced in [7]. They have an 
algebra analogous to that of the magnetic translations:
$$\hat t(n_1,n_1)\hat t(n_1',n_2')
=\exp\biggl[-{2\pi {\rm i}M\over{N}}(n_1n_2'-n_2n_1')\biggr]
\hat t(n_1+n_1',n_2+n_2')
\ .\eqno(8.2)$$
Using the definition (8.1), the coefficients $I_{nm}$
given by (7.17) which define the Dirac bracket (7.1) become
$$I_{n_1n_2}=(-1)^{pqn_1n_2}\sum_{\mu=0}^{\vert N\vert -1}
\bra{\phi'_\mu}\hat t(-qn_1,qn_2)
\hat T\bigl({n_2\over N}{\bf A}_1+{2\pi n_1\over \kappa}{\bf A}_2\bigr)
\ket {\phi_\mu}
\ .\eqno(8.3)$$
A further simplification can be introduced by using
the notation $\vert \Phi )$ to represent the set of
$N$ Wannier state vectors $\{\ket{\phi_\mu},\ \mu=0,..,\vert N\vert -1\}$.
The object $\vert \Phi )$ may be thought of as a state
vector in an \lq expanded' Hilbert space, with inner product
$$(\Phi'\vert\Phi)=\sum_{\mu=0}^{\vert N\vert -1}
\braket {\phi'_\mu}{\phi_\mu}
\ .\eqno(8.4)$$
Equation (8.3) can now be reduced to a satisfyingly simple
form by introducing a generalised magnetic translation operator
in the expanded Hilbert space:
$$\hat {\cal T}({\bf R})=(-1)^{pqn_1n_2}\hat t(-qn_1,qn_2)
\hat T\bigl({n_2\over N}{\bf A}_1
+{2\pi n_1\over \kappa}{\bf A}_2\bigr)
\ ,\ \ \ 
{\bf R}=n_1{\bf A}_1+n_2{\bf A}_2
\ .\eqno(8.5)$$
With this definition
$$I_{n_1n_2}=(\Phi'\vert \hat {\cal T}({\bf R})\vert \Phi)
\ .\eqno(8.6)$$
Also, comparing with (5.9), it can be that the gauge transformation
$\exp[{\rm i}{\bf k}.{\bf R}^\ast]$ results in a transformation of the
vector of type II Wannier states of the form
$$\vert \Phi')=\hat {\cal T}({\bf R}^\ast )\vert \Phi )
\ .\eqno(8.7)$$
The operators ${\cal T}({\bf R})$ again have a non-commuting
algebra analogous to that of the magnetic translations:
$$\hat {\cal T}({\bf R})\hat {\cal T}({\bf R}')=
\exp\biggl[\pi {\rm i}\gamma 
{({\bf R}\wedge {\bf R}')\over{({\bf A}_1\wedge{\bf A}_2)}}\biggr]
\hat {\cal T}({\bf R}+{\bf R}')
\eqno(8.8)$$
where
$$\gamma ={\Delta k\over \kappa}={q\beta-p\over{M+\beta N}}
\eqno(8.9)$$ 
is the dimensionless magnetic field parameter mentioned in the
introduction (equation (1.5)).

From (3.16), it is seen that the evaluation of the matrix elements 
of the Hamiltonian involves calculating the matrix elements
$\bra {B({\bf k}')}\hat T({\bf r})\ket {B({\bf k})}$, where
${\bf r}=\beta (\nu_1{\bf A}_1+\nu_2{\bf A}_2)$. The Dirac
bracket $\braket {B'({\bf k}')}{B({\bf k})}$ was obtained in
equations (7.16) and (7.17) in terms of a set of coefficients
$I_{n_1n_2}$. The calculation of section 6 shows that the operator
$\hat T({\bf r})$ acting on a Bloch state creates a new canonical
Bloch state with ${\bf k}$ shifted to ${\bf k}+\nu_2{\bf a}_1$.
It is natural to expect that the Dirac bracket 
$\braket{B'({\bf k}')}{B({\bf k}+\nu_2{\bf a}_1)}$ may be
expressed in the form (7.16), with the coefficients $I_{n_1n_2}$
replaced by $I_{\Phi'\Phi}(n_1,n_2,\nu_2)$ (note that 
$I_{n_1n_2}=I_{\Phi'\Phi}(n_1,n_2,0)$). Noting that
$p\kappa +M\Delta k=2\pi$, and that $k_1$ is replaced by
$k_1+2\pi \nu_2$ in (7.16), it is found that 
$$\braket {B'({\bf k}')}{B({\bf k}+\nu_2{\bf a}_1)}=
\sum_{N_1=-\infty}^\infty \sum_{N_2=-\infty}^\infty
\sum_{n_1=-\infty}^\infty \sum_{n_2=-\infty}^\infty
(-1)^{N_1n_2+pN_2n_1+MN_1N_2}$$
$$\times\delta(k_2-k_2'-2\pi N_2) 
\delta (k_1-k_1'-N_1\kappa-n_1\Delta k)
\exp\biggl[{\rm i}\biggl({k_2+k_2'\over{2}}\biggr)(pn_1-MN_1)\biggr]$$
$$\times 
\exp\biggl[-2\pi {\rm i}\biggl({k_1+k_1'\over{2\kappa}}\biggr)n_2\biggr]
I_{\Phi'\Phi}(n_1,n_2,\nu_2)
\eqno(8.10)$$
where
$$I_{\Phi'\Phi}(n_1,n_2,\nu_2)=\exp[{\rm i}\pi (p+2\pi /\kappa)n_2\nu_2]
I_{n_1+M\nu_2,n_2}
\eqno(8.11)$$
Now consider the evaluation of the matrix element
$\bra{B'({\bf k}')}\hat T({\bf r})\ket {B({\bf k})}$. This may 
be written in the form (8.10), with the Wannier state $\vert \Phi)$
replaced by the state $\vert \Phi'')$ given by (6.8). Combining
(6.8) and (8.11), the coefficients
may be written in the form (8.10) with coefficients
$$I_{\Phi'\Phi''}(n_1,n_2,\nu_1,\nu_2)=
\exp[{\rm i}\pi (p+2\pi /\kappa)n_2\nu_2]
\sum_{\mu=0}^{\vert N\vert -1}
\exp\biggl[{2\pi {\rm i}M\mu \nu_1\over N}\biggr]$$
$$\times
\bra {\phi'_\mu}
\hat {\cal T}({\bf R}+M\nu_2{\bf A}_1)
\hat T\biggl({\kappa\beta\over{2\pi N}}\nu_1{\bf A}_1
+{2\pi N\beta\over \kappa}\nu_2{\bf A}_2\biggr)\ket {\phi_\mu }
\ .\eqno(8.12)$$
This coefficient may be expressed in the form
$$
I_{\Phi'\Phi''}(n_1,n_2,\nu_1,\nu_2)=(\Phi'\vert \hat {\cal T}({\bf R})
\hat \tau ({\bf r})\vert \Phi )
\eqno(8.13)$$
where 
$$\hat \tau ({\bf r})=
\hat t(-\nu_2,\nu_1)\hat T\biggl({\kappa\beta\over{2\pi N}}\nu_1{\bf A}_1
+\beta \nu_2{\bf A}_2\biggr)\ .\eqno(8.14)$$
The operators $\cal \tau ({\bf r})$ commute with the 
$\hat {\cal T}({\bf R})$ operators:
$$\bigl[\hat \tau ({\bf r}),\hat {\cal T}({\bf R})\bigr]=0
\eqno(8.15)$$
for all lattice vectors ${\bf R}$ and ${\bf r}/\beta$.
\gap\gap\gap
%
%
%
%
\noindent{\bf 9. The generalised Peierls effective Hamiltonian}
\gap
\noindent{\sl 9.1 A one dimensional effective Hamiltonian}
\gap
The motivation is to obtain an effective Hamiltonian, having
a spectrum which is the same as a subset of the spectrum 
of the original Hamiltonian. The effective Hamiltonian is easier
to analyse because the number of degrees of freedom has been reduced.
The approach is analogous to that used in earlier work on the
phase space lattice Hamiltonian [7,14]. The Hamiltonian will be reduced 
to a block diagonal form,
and matrix elements of the Hamiltonian within one block will be
compared with
matrix elements of the effective Hamiltonian. If the basis states
are in one to one correspondence and the matrix elements are equal, 
then the spectrum of the effective Hamiltonian is the same as that
of the block of the full Hamiltonian.

In the case under consideration, matrix elements of the Hamiltonian
will be evaluated in the basis formed by a set of generalised Bloch 
states $\ket {B'({\bf k})}$. They are compared with 
matrix elements of an effective Hamiltonian $\hat H_{\rm proj}$ 
in a suitable basis with elements $\ket {\bar \xi(x,k_2)}$, and 
the coefficients 
defining $\hat H_{\rm proj}$ are chosen such that the non-zero
matrix elements of $\hat H_{\rm proj}$ correspond with those of
$\hat H$, in that
$$\bra{B({\bf k}')}\hat H\ket{B({\bf k})}
={4\pi^2\over {N\kappa}}\bra {\bar \xi (x',k_2)}\hat H_{\rm proj}
\ket {\bar \xi (x,k_2)}
\delta (k_2-k_2')
\ .\eqno(9.1)$$
This equality holds when $k_2=k_2'$, and where the states
$\ket {\bar \xi(x,k_2)}$ are labelled by a continuous variable 
$x= k_1/\kappa $.

It will be assumed that in the case where the 
dimensionless magnetic field $\beta$ takes the 
rational value $p/q$, there is a non-degenerate
band. It will also be assumed that the gap separating
this band from the rest of the spectrum does not
close when $\beta$ is perturbed away from the rational
value $p/q$. 
The effective Hamiltonian is constructed to reproduce
that part of the full spectrum which evolves out this
band when $\beta $ is perturbed from the rational value.
The type II Wannier functions $\ket{\phi_\mu}$ for 
this band are determined, and used to generate a set
of generalised Bloch states using (4.1). 
A projection operator $\hat P=f(\hat H)$ is 
applied to these states, where the function $f(E)$ is unity
where $E$ lies inside the band, and zero throughout the rest
of the spectrum. The states resulting from applying
this projection
$$\ket{B'({\bf k})}=\hat P \ket {B({\bf k})}\eqno(9.2)$$
are orthogonal to all eigenstates outside the band, and therefore
represent the Hamiltonian in block diagonal form.
The projection operator may be written in the form
$$\hat P=\int_{-\infty}^\infty dt\, \tilde f(t)\, \exp [{\rm i}\hat H t]
\eqno(9.3)$$
where $\tilde f(t)$ is a Fourier transform of $f(E)$.
The stipulation that the spectrum has a gap
on either side of the band ensures that $f(E)$ can have arbitrarily
many continuous derivatives, implying that this integral is
nicely behaved.

The projected Bloch generalised Bloch states are sufficiently 
numerous to form a complete but not overcomplete 
set for the band, and may be assumed to be complete provided
the matrix element $\braket{B'({\bf k}')}{B'({\bf k})}$
is sufficiently small when ${\bf k}\ne{\bf k}'$. This 
criterion can be tested and verified using the results of sections
7 and 8.
Because the states are not orthonormal, a normalisation
operator must also be calculated, such that
$$\braket{B'({\bf k}')}{B'({\bf k})}
={4\pi^2\over{N\kappa}}\bra {\bar \xi(x',k_2)}\hat N_{\rm proj}
\ket {\bar \xi(x,k_2)}
\delta (k_2-k_2')
\ .\eqno(9.4)$$
The subset of the spectrum of the full Hamiltonian
which lies in the projected band can be determined exactly
by solving the eigenvalue problem 
$\bigl[\hat H_{\rm proj}-E\hat N_{\rm proj}\bigr]\ket{\psi}=0$,
or alternatively by calculating the spectrum of the 
effective Hamiltonian operator
$$\hat H_{\rm eff}=\hat N_{\rm proj}^{-1/2}\hat H_{\rm proj}
\hat N_{\rm proj}^{-1/2}
\ .\eqno(9.5)$$
\par
Consider the matrix elements of the Hamiltonian, expressed
in the form (3.16), in the basis formed by the generalised 
Bloch states. 
The wavevectors ${\bf k}$ and ${\bf k}'$ can both be
restricted to the first Brillouin zone, i.e. 
$k_1,k_1'\in [0,\kappa)$ and $k_2,k_2'\in [0,2\pi)$,
because these states form a complete set. Alternatively,
states in an extended Brillouin zone can be used, since
they only differ by a phase factor from the states within 
the first Brillouin zone. States with $k_1$ differing by
multiples of $\kappa $ are identical (apart from a phase 
factor). Similiarly, states with $k_2$ differing by multiples
of $2\pi $ are identical. When writing matrix elements of the
Hamiltonian in a complete set of states, the summations over 
$N_1$ and $N_2$ in (7.16) can therefore be dropped: 
$$\bra {B({\bf k}')}\hat H\ket {B({\bf k})}
={4\pi^2\over N}\delta (k_2-k_2')\sum_{n_1=-\infty}^\infty
\delta(k_1-k_1'-n_1\Delta k)$$
$$\times\exp[{\rm i}pk_2n_1]\sum_{n_2=-\infty}^\infty
\exp\biggl[-2\pi{\rm i}\biggl({k_1+k_1'\over{2\kappa}}\biggr)n_2\biggr]
H'_{n_1n_2}
\ .\eqno(9.6)$$
In the case where $\beta $
is rational, $n\kappa +m\Delta k=0$ for some choice of
$n$ and $m$. In particular, $\gamma =\Delta k/\kappa$ is
also a rational number, $\gamma=p'/q'$, so that this 
relationship is satisfied when n is a multiple of $q'$.
In this case, only $q'$ distinct states are coupled, and 
the Hamiltonian is represented by a $q'\times q'$ matrix
with parameters $k_2\in [0,2\pi)$ and $k_1\in [0,\kappa/q')$.
In the general case there is no finite dimensional
representation.

Now compare the matrix elements (9.6) with matrix elements 
of an effective Hamiltonian of the form
$$\hat H_{\rm proj}=H_{\rm proj}(\hat {\bf K})
=\sum_{\bf R} H'({\bf R})\ \exp[{\rm i}\hat {\bf K}.{\bf R}]
\equiv \sum_{\bf R} H'({\bf R})\ \hat T'({\bf R})
\eqno(9.7)$$
where sum runs over all of the lattice vectors 
${\bf R}=n_1{\bf A}_1+n_2{\bf A}_2$, and where the following 
relations hold:
$$\hat {\bf K}={1\over {2\pi}}({\bf a}_1\hat g_1+{\bf a}_2\hat g_2)
\eqno(9.8)$$
$$[\hat g_1,\hat g_2]=2\pi {\rm i}\gamma
\eqno(9.9)$$
(here the ${\bf a}_i$ are reciprocal lattice vectors, satisfying
${\bf a}_i.{\bf A}_j=2\pi \delta_{ij}$).
The operators $\hat g_1$ and $\hat g_2$ have a commutator which is 
analogous to the usual position and momentum operators. Eigenstates
of $\hat g_2$ will be introduced, with eigenvalue $x$:
$\hat g_2 \ket {\xi(x)}=x\ket {\xi (x)}$.
Evaluating the matrix elements of (9.7) in this basis leads
to matrix elements which are very similar in structure to (9.6),
if the coefficients $H'({\bf R})$ in (9.7) are identified with
the coefficients $H'_{n_1n_2}$ in (9.6). The correspondence
becomes even closer if the states $\ket {\xi(x)}$ are
\lq gauge-transformed' as follows: 
$$\ket {\bar \xi(x,k_2)}=\exp
\biggl[{\rm i}\biggl({pk_2\over {2\pi \gamma}}\biggr)x\biggr]
\ket {\xi(x)}
\ .\eqno(9.10)$$
The matrix elements are then
$$\bra {\bar\xi(x',k_2)}\hat H_{\rm proj}\ket {\bar\xi(x,k_2)}=
\sum_{n_1=-\infty}^\infty \sum_{n_2=-\infty}^\infty
H'({\bf R})\exp[{\rm i}(x+x')n_2/2]$$
$$\times
\exp[{\rm i}pn_1k_2]
\delta (x-x'-2\pi \gamma n_1)
\ .\eqno(9.11)$$
Identifying $x=k_1/\kappa$ and $\gamma=\Delta k/\kappa$, these 
matrix elements of $\hat H_{\rm proj}$ are identical to the 
elements (9.6) for all values of $k_2$. The spectrum of (9.7) 
is therefore identical to that of (9.6) when $\gamma=\Delta k/\kappa$
and $H'({\bf R})=H'_{n_1n_2}$.
\gap
\noindent {\sl Coefficients of the effective Hamiltonian}
\gap
It remains to determine the coefficients $H'({\bf R})=H'_{n_1n_2}$ 
in (9.6).
These are obtained using (7.16) and the notational devices
introduced in section 8. The Hamiltonian
is given by (3.16), and takes the form of a sum of magnetic
translations of the form $\hat T({\bf r})$, where 
${\bf r}/\beta$ are lattice vectors. The action of the Hamiltonian
(3.16) upon a Bloch state $\ket {B({\bf k})}$ may be represented
in terms of the action of an image Hamiltonian upon the Wannier states
that generate the Bloch states. The matrix elements of the Hamiltonian,
$\bra {B({\bf k}')}\hat H\ket {B({\bf k})}$, are of the form (8.10),
with the coefficients $I_{\Phi'\Phi}(n_1,n_2,\nu_2)$ replaced by coefficients 
$H'_{n_1n_2}=H'({\bf R})$ characterising the Hamiltonian.
These are given by an expression analogous
to (8.13):
$$H'({\bf R})=(\Phi \vert \hat {\cal T}({\bf R})\,\hat {\cal H}\vert \Phi )
\ .\eqno(9.12)$$
The operators $\hat V_{\bf k}$
in (3.16) commute with the magnetic translations, and therefore
commute with $\hat \tau ({\bf r})$ and $\hat {\cal T}({\bf R})$.
Using (8.13) and (3.16), it is seen that the operator
$\hat {\cal H}$, which is the image of the Hamiltonian
in the Wannier function Hilbert space, is
$$\hat {\cal H}=\sum_{\bf k} \hat V_{\bf k}\ 
\hat \tau \bigl({\bf r}({\bf k})\bigr)
\eqno(9.13)$$
where ${\bf r}({\bf k})=\beta (n_2{\bf A}_1-n_1{\bf A}_2)$ 
corresponds to the reciprocal lattice vector 
${\bf k}=n_1{\bf a}_1+n_2{\bf a}_2$.
The image $\hat {\cal H}$ of Hamiltonian in the space of the 
Wannier states commutes with the image of the lattice
translation operators:
$$[\hat {\cal H},\hat {\cal T}({\bf R})]=0
\ .\eqno(9.14)$$
\par
A similar representation exists for the projection operator
$\hat P=f(\hat H)$: this has an image in the form of an 
operator $\hat {\cal P}$ acting upon the Wannier states.
Also, the 
operator $\hat {\cal H}_{\rm proj}=\hat {\cal P}\hat {\cal H}\hat {\cal P}$
which is the image of the projected Hamiltonian $\hat H_{\rm proj}$
acting on the Wannier functions may also be expressed in a
form analogous to (9.13). The effective Hamiltonian can also
be represented by an operator 
$\hat {\cal H}_{\rm eff}=\hat {\cal P}^{-1/2}\hat {\cal H}
\hat {\cal P}^{-1/2}$ acting on the Wannier states.

The formulae discussed above can be used to calculate the
Fourier coefficients of the effective Hamiltonian using
(9.12). Methods for calculating these coefficients as an
expansion in $\beta -p/q$ are discussed in [14] for the case
of the phase space lattice Hamiltonian, and these techniques
may be adapted to the present problem. In order to establish
the validity of the Peierls formula, it is necessary only
to establish the limit of the coefficients $H'({\bf R})$
in the limit $\beta \to p/q$. These coefficients are identified 
by noting that, upon setting $\beta =p/q$ the Bloch states become
eigenstates, so that 
$\bra{B({\bf k}')}\hat H\ket{B({\bf k})}={\cal E}({\bf k})
\delta ({\bf k}-{\bf k}')$. The corresponding expression (9.11) for 
the matrix elements of the effective Hamiltonian reduces to 
$$\bra {\bar \xi(x',k_2)}\hat H_{\rm eff}\ket {\bar \xi(x,k_2)}=
\sum_{n_1=-\infty}^\infty \sum_{n_2=-\infty}^\infty 
H'({\bf R})\delta (x-x')
\exp[{\rm i}xn_2]\exp[-{\rm i}pn_1k_2]
\ .\eqno(9.15)$$
In the limit $\beta \to p/q$, the coefficients $H'({\bf R})$ 
of the effective 
Hamiltonian are therefore the Fourier coefficients of the
dispersion relation. The effective Hamiltonian (9.7) is therefore
of the \lq Peierls substitution' form, (1.1).
\gap\gap\gap
%
%
%
%
\noindent{\bf 10. Acknowledgements}
\gap
This work was supported by the EPSRC, research grant 
reference GR/L02302.
\vfill
\eject
%
%
%
%
%
{\indentoff \bf References}
\gap
\indentoff
\gap
\ref {1}{G. H. Wannier}{Phys. Rev.}{52}{191}{1937}
[2] M. Ya. Azbel, {\sl Zh. eksp. teor. Fiz.}, {\bf 46}, 929, (1964).
({\sl transl. Sov. Phys. JETP}, {\bf 19}, 634-45, (1964)).
\gap
\ref {3}{D. R. Hofstadter}{Phys. Rev.}{B14}{2239-49}{1976}
\ref {4}{D. J. Thouless}{J. Phys.}{C17}{325-8}{1984}
\ref {5}{D. J. Thouless, M. Kohmoto, M. P. Nightingale,
and M. den Nijs}{Phys. Rev. Lett.}{49}{405-8}{1982}
\gap
\ref {6}{M. Wilkinson}{J. Phys.: Condensed Matter}{10}{7407}{1998}
\ref {7}{M. Wilkinson}{J. Phys.}{A27}{8123-48}{1994}
\ref {8}{M. Wilkinson and R. J. Kay}{J. Phys.}{A30}{5551}{1997}
\ref {9}{R. E. Peierls}{Z. Phys.}{80}{763-91}{1933}
\ref {10}{P. G. Harper}{Proc. Phys. Soc.}{A68}{879-92}{1955}
\ref {11}{R. Brown}{Phys. Rev.}{A133}{1038}{1964}
\ref {12}{J. Zak}{Phys. Rev.}{A134}{1602}{1964}
\ref {13}{M. C. Chang and Q. Niu}{Phys. Rev.}{B53}{7010}{1996}
\ref {14}{R. J. Kay and M. Wilkinson}{J. Phys.}{A29}{1515}{1996}
\ref {15}{P. St\v reda}{J. Phys.}{C15}{L717-27}{1982}
\vfill
\eject
\end